# Approximate Edge Analytics for the IoT Ecosystem

Zhenyu Wen*, Do Le Quoc†, Pramod Bhatotia*, Ruichuan Chen‡, Myungjin Lee*
*University of Edinburgh    †TU Dresden    ‡Nokia Bell Labs

*Abstract*—IoT-enabled devices continue to generate a massive amount of data. Transforming this continuously arriving raw data into timely insights is critical for many modern online services. For such settings, the traditional form of data analytics over the entire dataset would be prohibitively limiting and expensive for supporting real-time stream analytics.

In this work, we make a case for approximate computing for data analytics in IoT settings. Approximate computing aims for efficient execution of workflows where an approximate output is sufficient instead of the exact output. The idea behind approximate computing is to compute over a representative sample instead of the entire input dataset. Thus, approximate computing — based on the chosen sample size — can make a systematic trade-off between the output accuracy and computation efficiency.

This motivated the design of APPROXIOT— a data analytics system for approximate computing in IoT. To realize this idea, we designed an online hierarchical stratified reservoir sampling algorithm that uses edge computing resources to produce approximate output with rigorous error bounds. To showcase the effectiveness of our algorithm, we implemented APPROXIOT based on Apache Kafka and evaluated its effectiveness using a set of microbenchmarks and real-world case studies. Our results show that APPROXIOT achieves a speedup $1.3\times$—$9.9\times$ with varying sampling fraction of $80\%$ to $10\%$ compared to simple random sampling.

## I. INTRODUCTION

Most modern online services rely on timely data-driven insights for greater productivity, intelligent features, and higher revenues. In this context, the Internet of Things (IoT) — all of the people and things connected to the Internet — would provide important benefits for modern online services. IoT is expected to generate 508 zettabytes of data by 2019 with billions of new smart sensors and devices [1]. Large-scale data management and analytics on such "Big Data" will be a massive challenge for organizations.

In the current deployments, most of this data management and analysis is performed in the cloud or enterprise datacenters [2]. In particular, most organizations continuously collect the data in a centralized datacenter, and employ a stream processing system to transform the continuously arriving raw data stream into useful insights. These systems target low-latency execution environments with strict service-level agreements (SLAs) for processing the input data stream.

Traditionally, the low-latency requirement is usually achieved by employing more computing resources and parallelizing the application logic over the datacenter infrastructure. Since most stream processing systems adopt a data-parallel programming model such as MapReduce, almost linear scalability can be achieved with increased computing resources. However, this scalability comes at the cost of ineffective utilization of computing resources and reduced throughput of the system. Moreover, in some cases, processing the entire input data stream would require more than the available computing resources to meet the desired latency/throughput guarantees. In the context of IoT, transferring, managing, and analyzing large amounts of data in a centralized enterprise datacenter would be prohibitively expensive [3].

In this paper, we aim to build a stream analytics system to strike a balance between the two desirable but contradictory design requirements, i.e., achieving low latency for real-time analytics, and efficient utilization of computing resources. To achieve our goal, we propose a system design based on *approximate computing* paradigm that explores a novel design point to resolve this tension. In particular, approximate computing is based on the observation that many data analytics jobs are amenable to an approximate rather than the exact output [4], [5]. For such workflows, it is possible to trade the output accuracy by computing over a subset instead of the entire data stream. Since computing over a subset of input requires less time and computing resources, approximate computing can achieve desirable latency and computing resource utilization.

Furthermore, the heterogeneous edge computing resources have limited computational power, network bandwidth, storage capacity, and energy constraints [3]. To overcome these limitations, the approximate computing can be adapted to the available resources through trading off the accuracy and performance, while building a "truly" distributed data analytics system over IoT infrastructures such as mobile phones, PCs, sensors, network gateways/middleboxes, CDNs, and edge datacenters at ISPs.

We design and implement APPROXIOT to realize our vision for a low-latency and resource-efficient stream analytics system based on the above key observations. APPROXIOT recruits the aforementioned edge computing nodes and creates a stream processing pipeline as a logical tree (Figure 1). A data stream traverses over the logical tree towards a centralized cloud or datacenter where the data analysis queries are executed. Along the route to the central location, each node independently selects data items from the input stream while preserving statistical characteristics. The core of APPROXIOT's design is a novel online sampling algorithm that updates the significance (weight) of those selected data items on each node without any cross-node synchronization. The system can tune the degree of sampling systematically, depending on resource availability and analytics requirements.

Overall, this paper makes the following key contributions.

- **Approximate computing for IoT-driven stream analytics.** We make a case for approximate computing in IoT, whereby the real-time analysis over the entire data stream is becoming unsustainable due to the gap between the required computing resources and the data volume.
- **Design and implementation of APPROXIOT (§III and §IV).** We design the core algorithm of APPROXIOT—

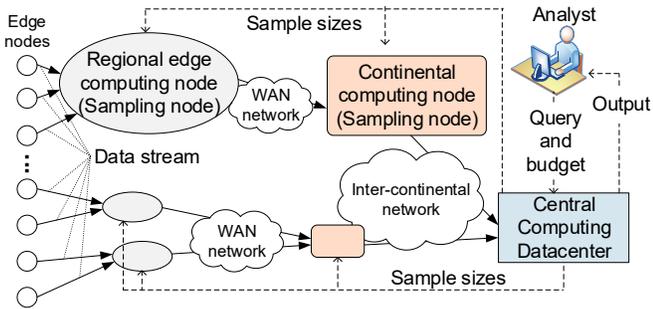

Fig. 1. System overview.

weighted hierarchical sampling — based on theoretical foundations. The algorithm needs no coordination across nodes in the system, thereby making APPROXIOT easily parallelizable and hence scalable. Moreover, our algorithm is suitable to process different types of input streams such as long-tailed streams and uniform-speed streams. We prototype APPROXIOT using Apache Kafka.
- **Comprehensive evaluation of APPROXIOT (§V and §VI).** We evaluate APPROXIOT with synthetic and real-world datasets. Our evaluation results demonstrate that APPROXIOT outperforms the existing approaches. It achieves $1.3\times$—$9.9\times$ higher throughput than the native stream analytics execution, and $3.3\times$—$8.8\times$ higher accuracy compared to a simple random sampling scheme.

## II. OVERVIEW AND BACKGROUND

### A. System Overview

APPROXIOT builds on two design concepts: hierarchical processing and approximate computing. In APPROXIOT, a wide variety of devices or sensors (so-called IoT devices) generate and send data streams to regional edge computing nodes geographically close to themselves. The edge computing clusters managed by local ISPs or content providers sample only a subset of the input data streams and forward them to larger computing facilities such as datacenters. The data streams, again sampled at the datacenters, can be further forwarded to a central location, where a user-specified query is executed and the query results are produced for global-level analysis. These computing clusters spread across the globe form a logical stream processing pipeline as a tree, which is collectively called APPROXIOT. Figure 1 presents the high-level structure of the system.

The design choice of APPROXIOT, i.e., combining approximate computing and hierarchical processing, naturally enables the processing of the input data stream within a specified resource budget. On top of this feature, APPROXIOT produces an approximate query result with rigorous error bounds. In particular, APPROXIOT designs a parallelizable online sampling technique to select and process a subset of data items, where the sample size can be determined based on the resource constraints at each node (i.e., computing cluster), without any cross-node coordination.

Altogether, APPROXIOT achieves three goals.

- **Resource efficiency.** APPROXIOT utilizes computing and bandwidth resources efficiently by sampling data items at each individual node in the logical tree. If we were to sample data items only at a node where the query is executed, all the computing and bandwidth resources used to process and forward the unused data items would have been wasted.
- **Adaptability.** The system can adjust the degree of sampling based on resource constraints of the nodes. While the core design is agnostic to the ways of choosing the sample size, i.e., whether it is centralized or distributed, this adaptability ensures better resource utilization.
- **Transparency.** For an analyst, the system enables computation over the distributed data in a completely transparent fashion. The analyst does not have to manage computational resources; neither does she require any code changes to existing data analytics application/query.

### B. Technical Building Blocks

APPROXIOT relies on two sampling techniques as the building blocks: stratified sampling [6] and reservoir sampling [7] because the properties of the two allow APPROXIOT to meet its needs.

*1) Stratified Sampling:* A sub-stream is the data items from a source. In reality, sub-streams from different data sources may follow different distributions. Stratified sampling was proposed to sample such sub-streams fairly. Here, each sub-stream forms a stratum; if multiple sub-streams follow the same data distribution, they can be combined to form a stratum. For clarity and coherence, hereafter, we still use sub-stream to refer to a stratum.

Stratified sampling receives sub-streams from diverse data sources, and performs the sampling (e.g., simple random sampling [8] or other types of sampling) over each sub-stream independently. In doing so, the data items from each sub-stream can be fairly selected into the sample. Stratified sampling reduces sampling error and improves the precision of the sample. It, however, works only in a situation where it can assume the knowledge of the statistics of all sub-streams (e.g., each sub-stream's length). This assumption on prior knowledge is unrealistic in practice.

*2) Reservoir Sampling:* Reservoir sampling is often used to address the unrealistic assumption aforementioned in stratified sampling. It works without the prior knowledge of all the sub-streams. Suppose a system receives a stream consisting of an unknown number of data items. Reservoir sampling maintains a reservoir of size $R$, and wants to select a sample of (at most) $R$ items uniformly at random from the unbounded data stream. Specifically, reservoir sampling keeps the first-received $R$ items in the reservoir. Afterwards, whenever the $i$-th item arrives ($i > R$), reservoir sampling keeps this item with probability of $N/i$ and then randomly replaces one existing item in the reservoir. In doing so, each data item in the unbounded stream is selected into the reservoir with equal probability. Reservoir sampling is resource-efficient; however, it could mutilate the statistical quality of the sampled data

## Algorithm 1: : APPROXIOT's algorithm overview

**Input**:
*query*: streaming query (only for root)
*budget*: resource budget to execute the query
*parent*: successor node

```
1  begin
2      foreach time interval do
3          size ← costFunction(budget)
4          // W^in: Weight set from downstream nodes
5          // C^in: Count set from downstream nodes
6          {W^in, C^in, items} ← getDataStream(interval)
7          // Weighted Hierarchical Sampling (§III-B)
8          // W^out: a map of weights of the sample
9          // C^out: a map of counts of the sample
10         {sample, W^out, C^out} ← WHSamp(items, size, W^in, C^in)
11         if parent is not empty then
12             // (weight, count, sample) to upstream node
13             Send(parent, W^out, C^out, sample)
14         end
15         else
16             // Run query as a data-parallel job
17             result ← runJob(query, sample, W^out)
18             // Estimate error bounds (§III-D)
19             error ← estimateError(result)
20             write result ± error
21         end
22     end
23 end
```

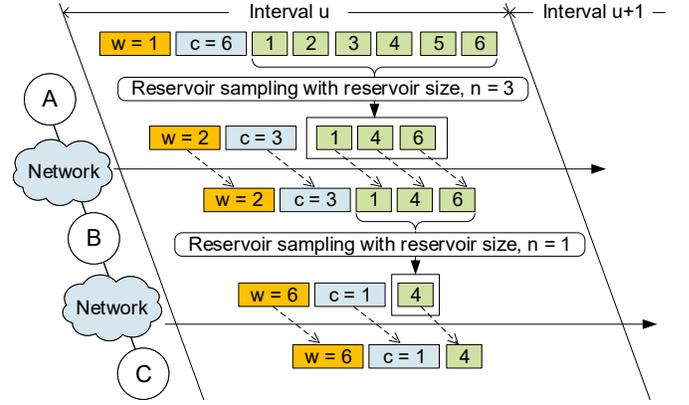

Fig. 2. Synchronized arrival of data stream. A node (e.g., $A$) receives sub-streams (only one sub-stream is shown for brevity) and an interval of a sub-stream contains weight ($w$), count ($c$), and a series of items. In this model, the count is equal to the number of items arriving in that interval. After reservoir sampling is applied, $w$ and $c$ are updated based on Algorithm 2. For example, node $A$ samples three out of 6 items; thus, $w = 2$ and $c = 3$.

items in the reservoir especially when the input data stream combines multiple sub-streams with different distributions. For example, the data items received from an infrequent sub-stream could easily get overlooked in reservoir sampling.

## III. DESIGN

In this section, we introduce the design of APPROXIOT. We first present its workflow (§III-A). Then, we detail its sampling mechanism (§III-B and §III-C), and error estimation mechanism (§III-D). Finally, we discuss design extensions to enhance the proposed system (§III-E).

### A. System Workflow

Algorithm 1 presents the overall workflow of APPROXIOT. The algorithm running at each node takes resource *budget* and *parent* as input while that of a root node additionally accepts a user-specified streaming *query*. A number of sources generate data items and continuously push them in a streaming fashion through a pre-configured logical tree. Each node in the tree samples data items on a sub-stream basis, based on a specified resource budget. Then, each node (denoted as sampling node in Figure 1) forwards those sampled sub-streams associated with a small amount of metadata to upstream node towards a root node. For sub-streams arriving at the root (aggregate node), the root conducts sampling of sub-streams, executes the query on the data items, and outputs query results alongside rigorous error bounds.

For each time interval, a node derives the sample size (*size*) based on the given resource budget (line 3). It then extracts data items and metadata—*weight set* and *count set*—for sub-streams that arrive within the interval (line 6). Next, it runs a weighted hierarchical sampling (WHSamp) using the items, weight set ($W^{in}$) and count set ($C^{in}$), and returns the sampled sub-streams, revised weight set, $W^{out}$ and count set, $C^{out}$ (line 10). If the node is a sampling node (if there is *parent* for the node), then the node sends the sample, $W^{out}$ and $C^{out}$ to its parent node (line 13). Otherwise, it processes the query on those data and as the last step, it runs an error estimation mechanism (as described in §III-D) to compute the error bounds for the approximate query result in the form of $output \pm error$ bound (lines 16-20).

The crux of APPROXIOT is the proposed hierarchical sampling algorithm (detailed in §III-B) that selects a portion of all sub-streams for the sample without neglecting any single sub-stream, for which we leverage and extend the existing stratified reservoir sampling [9].

The whole process repeats for each time interval as the computation window slides [10], [11]. Note that the resource budget can change across time intervals to adapt to user's requirements for the budget.

### B. Weighted Hierarchical Sampling

We assume that the items belonging to an interval arrive in a synchronized fashion. Figure 2 illustrates this synchronized arrival model where all the items arrive in the same interval when their associated sets $W^{in}$ and $C^{in}$ arrive. In §III-C, we relax this assumption.

Given this synchronization model, we design Algorithm 2. The algorithm outlines the high-level idea of the weighted hierarchical sampling in a node. The node first stratifies the input stream into sub-streams according to their sources (line 5). It then determines, denoted as $N_i$, the reservoir size for each sub-stream $S_i$ (line 7). Given the resovoir sizes, the node selects items at random from $S_i$ through the traditional reservoir sampling (line 10). The reservoir sampling ensures that the total number of selected items from $S_i$ does not exceed its sample size $N_i$. Given the input weight ($W_i^{in}$) for $S_i$, the

**Algorithm 2: : Weighted hierarchical sampling**

**Input**:
*items*: input data items
*sampleSize*: size of sample
$W^{in}$: weight set from downstream nodes
$C^{in}$: count set from downstream nodes

1 WHSamp(*items*, *sampleSize*, $W^{in}$, $C^{in}$)
2    // sample: set of items sampled within a time interval
3    sample ← ∅
4    // Update S, a set of sub-streams seen so far within the time interval
5    S ← Update(*items*)
6    // Decide the sample size for each sub-stream
7    N ← getSampleSize(*sampleSize*, S)
8    **forall** the $S_i \in S$ **do**
9       $c_i \leftarrow |S_i|$ // $S_i$: sub-stream i
10      $sample_i \leftarrow \text{RS}(S_i, N_i)$ // Reservoir sampling
11      // Compute the weight of $sample_i$ according to Equation 1
12      **if** $c_i > N_i$ **then**
13         $w_i \leftarrow \frac{c_i}{N_i}$
14         $W_i^{out} \leftarrow W_i^{in} * w_i$ // update weight of $S_i$
15         $C_i^{out} \leftarrow N_i$
16      **end**
17      **else**
18         $W_i^{out} \leftarrow W_i^{in}$
19         $C_i^{out} \leftarrow c_i$
20      **end**
21    **end**
22    **return** $W^{out}, C^{out}, sample$

---

node finally computes an effective weight (lines 12-20). This process repeats across all sub-streams. Finally, we return the final sample, weight set, and count set (line 22).

We now describe how to statistically recreate the original items from the sample and set of weights. In order to facilitate our discussion, we consider two cases: (i) single node and (ii) multiple nodes.

**Single node.** In this case, a node works as root. Initially, when a source generates data stream, there is no weight set given to the node, and each weight of input sub-streams is assumed to be 1 (i.e., $W_i^{in} = 1$). The reservoir sampling guarantees at most $N_i$ number of items is selected out of $c_i$ items from $S_i$. To reconstruct the statistics of the original items, we compute a specific weight ($w_i$) for the items selected from each sub-stream $S_i$ as follows:

$$w_i = \begin{cases} c_i/N_i & \text{if } c_i > N_i \\ 1 & \text{if } c_i \leq N_i \end{cases} \quad (1)$$

We finally calculate an effective weight for $S_i$ by multiplying $W_i^{in}$ by $w_i$, that is, $W_i^{out} = W_i^{in} * w_i$. Since $W_i^{in} = 1$, effectively $W_i^{out} = w_i$.

Given the sampled items and weight for each sub-stream, it is feasible to support linear queries with approximation. One such example is to compute the sum of all received items by computing an approximate weighted sum of sub-stream $S_i$. We denote the sum as $SUM_i$ and estimate it as follows:

$$SUM_i = (\sum_{k=1}^{Y_i} I_{i,k}) \cdot W_i^{out} \quad (2)$$

where $Y_i$ is the number of randomly selected items for $S_i$ ($Y_i \leq N_i$).

Suppose there are in total $X$ sub-streams $\{S_i\}_{i=1}^{X}$, the approximate total sum of all items received from all sub-streams (denoted as $SUM_*$) is:

$$SUM_* = \sum_{i=1}^{X} SUM_i \quad (3)$$

Note that Algorithm 2 works exactly the same way as one in [9] when there is only one node.

**Multiple nodes.** We extend our notations for further discussion. $SUM_{i,j}$ is an estimated sum of items in sub-stream $S_i$ at node $j$. We redefine $c_{i,j}$, $w_{i,j}$, $N_{i,j}$, $W_{i,j}^{out}$ and $Y_{i,j}$ in the similar fashion. We define $\mathcal{E}_{i,j}$ as a set of nodes on an *upstream path* to node $j$ (including $j$) for sub-stream $S_i$ from its original source. Also, let $\pi(i,j)$ be a predecessor node (i.e., an immediate downstream node) of node $j$ on an *upstream path* for sub-stream $S_i$. Then, we have:

$$SUM_{i,j} = (\sum_{k=1}^{Y_{i,j}} I_{i,j,k}) \cdot W_{i,j}^{out} \quad (4)$$

$$= (\sum_{k=1}^{Y_{i,j}} I_{i,j,k}) \cdot \max_{e \in \mathcal{E}_{i,j}} W_{i,e}^{out} \quad (5)$$

Here, $W_{i,j}^{out} = \max_{e \in \mathcal{E}_{i,j}} W_{i,e}^{out}$ because $W_{i,j}^{out} = W_{i,\pi(i,e)}^{out}$ as $w_{i,e} = 1$ if $c_{i,e} \leq N_{i,e}$ for any $e \in \mathcal{E}_{i,j}$ (see line 18 in Algorithm 2).

Now let $\chi(i,j)$ be an index of the first node that has the largest weight among nodes in $\mathcal{E}_{i,j}$. Then, $W_{i,j}^{out} = W_{i,\chi(i,j)}^{out} = \max_{e \in \mathcal{E}_{i,j}} W_{i,e}^{out}$. If $W_{i,j}^{out} = 1$, we consider that $\chi(i,j)$ does not exist. Given these basics, $W_{i,j}^{out}$ is expressed as follows.

$$W_{i,j}^{out} = \begin{cases} c_{i,src}/N_{i,\chi(i,j)} & \text{if } c_{i,src} > N_{i,\chi(i,j)} \\ 1 & \text{if } c_{i,src} \leq N_{i,k}, \forall k \in \mathcal{E}_{i,j} \end{cases} \quad (6)$$

Here, $c_{i,src}$ is the number of items for sub-stream $S_i$ at its original source. We can prove this by induction. For this discussion, let $v(i,j)$ be an immediate upstream node of node $j$ on an *upstream path* for sub-stream $S_i$.

**Base case:** $W_{i,v(i,src)}^{out} = c_{i,src}/N_{i,v(i,src)}$ if $c_{i,src} > N_{i,v(i,src)}$; $W_{i,v(i,src)}^{out} = 1$ otherwise.

Let $j$ denote $v(i,src)$. In this case, if $c_{i,src} > N_{i,j}$, $\chi(i,j) = j$. Thus, $W_{i,j}^{out} = W_{i,\chi(i,j)}^{out} = c_{i,src}/N_{i,\chi(i,j)}$. If $c_{i,src} \leq N_{i,j}$, obviously $W_{i,j}^{out} = 1$.

**Inductive step:** $W_{i,v(i,k)}^{out} = W_{i,k}^{out} \cdot c_{i,v(i,k)}/N_{i,v(i,k)}$ if $c_{i,v(i,k)} > N_{i,v(i,k)}$; $W_{i,v(i,k)}^{out} = W_{i,k}^{out}$ otherwise.

Let $j$ denote $v(i,k)$. We first have $W_{i,k}^{out} = W_{i,\chi(i,k)}^{out} = c_{i,src}/N_{i,\chi(i,k)}$. Then, $W_{i,j}^{out} = c_{i,src}/N_{i,\chi(i,k)} \cdot c_{i,j}/N_{i,j}$.

Suppose that $c_{i,j} > N_{i,j}$. By definition, $\chi(i,k)$ represents the first node that has the largest weight along an upstream path to node $k$ for sub-stream $S_i$. This means that any node $p \in \mathcal{E}_{i,k} \setminus \mathcal{E}_{i,\chi(i,k)}$ has $w_{i,p} = 1$. Therefore, $N_{i,\chi(i,k)} = c_{i,j}$ because $N_{i,\chi(i,k)}$ number of items sampled at node $\chi(i,k)$ arrive at node $j$ within the same time interval. Hence, $W_{i,j}^{out} = c_{i,src}/N_{i,j}$. Since $W_{i,j}^{out} > W_{i,k}^{out}$, $\chi(i,j) = j$. Therefore, $W_{i,j}^{out} = W_{i,\chi(i,j)}^{out} = c_{i,src}/N_{i,\chi(i,j)}$.

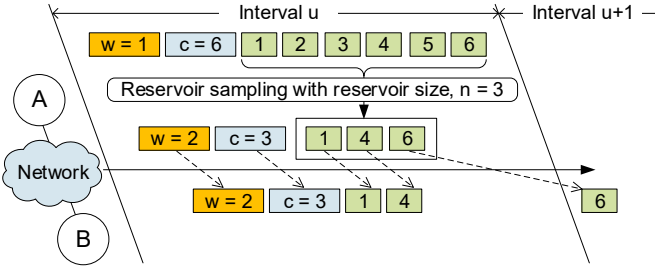

Fig. 3. Asynchronized arrival of data stream.

If there is no $\chi(i,k)$ that meets the inequality $c_{i,src} > N_{i,\chi(i,k)}$, $W_{i,k}^{out} = 1$. Then, $W_{i,j}^{out} = c_{i,j}/N_{i,j}$ if $c_{i,j} > N_{i,j}$. Otherwise, $W_{i,j}^{out} = 1$. This is equivalent to the base case. ∎

Now when $W_{i,j}^{out} > 1$, Equation (4) is reduced to:

$$SUM_{i,j} = (\sum_{k=1}^{Y_{i,j}} I_{i,j,k}) \cdot \frac{c_{i,src}}{N_{i,\chi(i,j)}} \quad (7)$$

Note that $Y_{i,j} = N_{i,\chi(i,j)}$ because node $\chi(i,j)$ is node $j$ or the weight is 1 for all nodes between node $\chi(i,j)$ and node $j$ along the upstream path.

Now Equation (4) meets the following property: $SUM_{i,p} \simeq SUM_{i,q}$, where $p, q \in \mathcal{E}_{i,j}$ ($p \neq q$). From Equation (7), we know that $SUM_{i,p} = (\sum_{k=1}^{Y_{i,p}} I_{i,p,k}) \cdot c_{i,src}/N_{i,\chi(i,p)}$ and $SUM_{i,q} = (\sum_{k=1}^{Y_{i,q}} I_{i,q,k}) \cdot c_{i,src}/N_{i,\chi(i,q)}$.

This quantity $\sum_{k=1}^{Y_{i,p}} I_{i,p,k}/N_{i,\chi(i,p)}$ is an average of sampled items for sub-stream $S_i$ since $Y_{i,p} = N_{i,\chi(i,p)}$. Since this query is a linear type, the two sample averages from nodes $p$ and $q$ should be similar. Hence, $SUM_{i,p} \simeq SUM_{i,q}$.

Therefore, even when multiple nodes do sampling independently, the root node can produce an undistorted query result on the input stream while the degree of error is bounded by the actual sampling rate, which we discuss in §III-D.

### C. Handling Asynchronous Time Interval

One assumption made for Algorithm 2 is that time intervals across nodes are synchronized and thus all sampled sub-streams and a set of weights associated with them from a downstream node arrive at an upstream node within a time interval. In other words, the sub-streams and set of weights that belonged to an interval at the downstream node do not straddle a time interval at the upstream node. This assumption is unlikely to hold true in practice due to various reasons such as varying bandwidth between nodes, different resource availability across nodes, and so forth. We now relax the assumption and let each node independently maintain intervals. To handle this issue of asynchronous time interval across nodes, we extend Algorithm 2 by taking into account the actual counts of sampled items for sub-streams.

For node $j \in \mathcal{E}_{i,j}$, suppose node $\pi(i,j)$ sends $W_{i,\pi(i,j)}^{out}$ to node $j$, followed by a series of items of sub-stream $S_i$ selected from node $\pi(i,j)$. Now let us consider the case in question, where the number of items arriving at node $j$ is less than that of sampled items at node $\pi(i,j)$. That is, $c_{i,j} = \alpha C_{i,\pi(i,j)}^{out}$ where $0 < \alpha < 1$ and $C_{i,\pi(i,j)}^{out}$ is the number of sampled items for $S_i$ at node $\pi(i,j)$, which is $\min(c_{i,\pi(i,j)}, N_{i,\pi(i,j)})$. When $c_{i,j} < N_{i,j}$, node $j$ selects all the items of sub-stream $S_i$. As such, there is no need for updating $W_{i,j}^{out}$ and an asynchronous time interval is not an issue at all. Thus, we only consider a case where $c_{i,j} > N_{i,j}$ and thus $\chi(i,j) = j$. In this case, Equation (4) becomes the following:

$$\begin{aligned}
SUM_{i,j} &= (\sum_{k=1}^{Y_{i,j}} I_{i,j,k}) \cdot W_{i,\pi(i,j)}^{out} \cdot \frac{c_{i,j}}{N_{i,j}} \\
&= (\sum_{k=1}^{Y_{i,j}} I_{i,j,k}) \cdot W_{i,\pi(i,j)}^{out} \cdot \frac{\alpha C_{i,\pi(i,j)}^{out}}{N_{i,j}} \\
&= (\sum_{k=1}^{Y_{i,j}} I_{i,j,k}) \cdot \frac{\alpha c_{i,src}}{N_{i,\chi(i,j)}}
\end{aligned} \quad (8)$$

Given Equation (8), $SUM_{i,j}$ is under-estimated by a factor of $\alpha$ ($= c_{i,j}/C_{i,\pi(i,j)}^{out}$). To correct this bias, we calibrate $W_{i,j}^{out}$ by $1/\alpha$. Hence, the calibrated weight $W_{i,j}^{out}$ is $W_{i,j}^{in} \times w_{i,j} \times C_{i,\pi(i,j)}^{out}/c_{i,j}$. This also suggests that the number of sampled items for sub-stream $i$ at the downstream node $\pi(i,j)$ should be known to node $j$. We use $C_{i,j}^{in}$ to denote $C_{i,\pi(i,j)}^{out}$; and $C_{i,j}^{in}$ is another input parameter fed to the algorithm along with $W_{i,j}^{in}$. The final equation is expressed as

$$W_{i,j}^{out} = W_{i,j}^{in} \times w_{i,j} \times C_{i,j}^{in}/c_{i,j}. \quad (9)$$

Therefore, Algorithm 2 is modified as follows. To correct the bias, we use the values in $C^{in}$; the expression on line 14 in Algorithm 2 should be replaced with Equation 9.

Now there is one remaining issue; in this asynchronous time interval case, some items do not have their associated $W^{in}$ and $C^{in}$ in their arriving interval (e.g., see item 6 in Figure 3). To handle the issue, each node maintains the most recent sets $W^{in}$ and $C^{in}$. Each value in these sets is only updated when a new value for a corresponding sub-stream arrives at a node. When the sampled items from those items are forwarded to an upstream node, new $W^{out}$ and $C^{out}$ sets are also generated and prepended to the items. For sub-stream $i$, $W_i^{in}$ and $C_i^{in}$ arrive from a downstream node, those values are updated in the locally-maintained sets $W^{in}$ and $C^{in}$ in Algorithm 1.

**Example.** Consider a case in Figure 4. Because intervals between nodes 1 and 2 are not aligned (see Figure 4(b)), a calibrated weight at node 2 should be readjusted. At node 1 in the example, $W_{i,1}^{out} = c_{i,src}/N_{i,1}$. At node 2, $\alpha = c_{i,2}/N_{i,1}$. Hence, $W_{i,2}^{out} = W_{i,1}^{out} \times c_{i,2}/N_{i,2}/\alpha = c_{i,src}/N_{i,1} \times c_{i,2}/N_{i,2} \times N_{i,1}/c_{i,2} = c_{i,src}/N_{i,2}$. Since $SUM_{i,1} = (\sum_{k=1}^{Y_{i,1}} I_{i,1,k}) \times c_{i,src}/N_{i,1}$ and $SUM_{i,2} = (\sum_{k=1}^{Y_{i,2}} I_{i,2,k}) \times c_{i,src}/N_{i,2}$, $SUM_{i,1} \simeq SUM_{i,2}$.

For $(1-\alpha)N_{i,1}$ items, node 2 uses the saved $W_{i,2}^{in}$ and $C_{i,2}^{in}$ where $W_{i,2}^{in} = W_{i,1}^{out}$ and $C_{i,2}^{in} = N_{i,1}$. Thus, the calibrated weight $W_{i,2}^{out} = c_{i,src}/N_{i,1} \times (1-\alpha)N_{i,1}/N_{i,2} \times C_{i,2}^{in}/((1-\alpha)N_{i,1}) = c_{i,src}/N_{i,2}$.

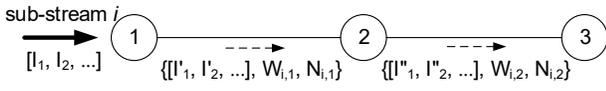

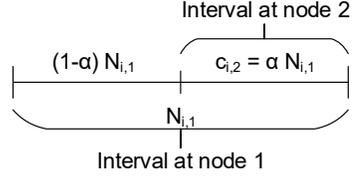

(a) Example topology

(b) Misaligned interval viewed at Node 2

Fig. 4. A simple example.

### D. Error Estimation

We described how we apply the proposed algorithm to randomly sample the input data stream to generate the approximate results for linear queries. We now describe a method to estimate the accuracy of our approximate results via rigorous error bounds.

As Algorithm 1 shows that the queries and error estimation will only perform on *root*, say node $j$. We assume that there are $X$ geo-distributed sub-streams $\{S_i\}_{i=1}^X$ as the input stream of APPROXIOT. We compute the approximate sum of all items received from all sub-streams through a logical aggregation tree, randomly sampling only $Y_{i,j}$ items from each sub-stream $S_{i,j}$. As each sub-stream is sampled independently, the variance of the approximate sum is:

$$Var(SUM_{*,j}) = \sum_{i=1}^{X} Var(SUM_{i,j}) \quad (10)$$

Further, as items are randomly selected for a sample within each sub-stream, according to the random sampling theory (Central limit theorem) [12], the variance of the approximate sum can be estimated as:

$$\widehat{Var}(SUM_{*,j}) = \sum_{i=1}^{X} \left( c_{i,src} \cdot (c_{i,src} - Y_{i,j}) \cdot \frac{s_{i,j}^2}{Y_{i,j}} \right) \quad (11)$$

We can easily compute $c_{i,src}$ by $Y_{i,j} \cdot W_{i,j}^{out}$ because either $Y_{i,j} = N_{i,\chi(i,j)}$ or $Y_{i,j} = c_{i,src}$. Here, $s_{i,j}$ denotes the standard deviation of the sub-stream $S_i$'s sampled items at node $j$:

$$s_{i,j}^2 = \frac{1}{Y_{i,j} - 1} \cdot \sum_{k=1}^{Y_{i,j}} (I_{i,j,k} - \bar{I}_{i,j})^2 \quad (12)$$

where $\bar{I}_{i,j} = \frac{1}{Y_{i,j}} \cdot \sum_{k=1}^{Y_{i,j}} I_{i,j,k}$.

Next, we show how we can also estimate the variance of the approximate mean value of all items received from all the $X$ sub-streams. The approximate mean value can be computed as:

$$MEAN_{*,j} = \frac{SUM_{*,j}}{\sum_{i=1}^{X} c_{i,src}} = \frac{\sum_{i=1}^{X} c_{i,src} \cdot MEAN_{i,j}}{\sum_{i=1}^{X} c_{i,src}}$$
$$= \sum_{i=1}^{X} (\varphi_i \cdot MEAN_{i,j}) \quad (13)$$

Here, $\varphi_i = \frac{c_{i,src}}{\sum_{i=1}^{X} c_{i,src}}$. Then, as each sub-stream is sampled independently, according to the random sampling theory [12], the variance of the approximate mean value can be estimated as:

$$\widehat{Var}(MEAN_{*,j}) = \sum_{i=1}^{X} Var(\varphi_i \cdot MEAN_{i,j})$$
$$= \sum_{i=1}^{X} \varphi_i^2 \cdot Var(MEAN_{i,j}) \quad (14)$$
$$= \sum_{i=1}^{X} \varphi_i^2 \cdot \frac{s_{i,j}^2}{Y_{i,j}} \cdot \frac{c_{i,src} - Y_{i,j}}{c_{i,src}}$$

Above, we have shown how to estimate the variances of the approximate sum and the approximate mean of the input data stream. Similarly, based on the central limit theorem, we can easily estimate the variance of the approximate results of any linear queries.

**Error bound.** We compute the error bound for the approximate result based on the '68-95-99.7" rule [13]. According to this rule, the approximate result is within one, two, and three standard deviations away from the exact result with probabilities of 68%, 95%, and 99.7%, respectively. The standard deviation is computed by by taking the square root of the variance in Equation 11 and Equation 14.

### E. Distributed Execution

Our proposed algorithm naturally extends for distributed execution as it does not require synchronization. Our straightforward design extension for parallelization is as follows: we handle each sub-stream by a set of $w$ worker nodes. Each worker node samples an equal portion of items from this sub-stream and generates a local reservoir of size no larger than $N_i/w$, where $N_i$ is the total reservoir size allocated for sub-stream $S_i$. In addition, each worker node maintains a local counter to measure the number of its received items within a concerned time interval for weight calculation. The rest of the design remains the same.

## IV. IMPLEMENTATION

We implemented APPROXIOT using Apache Kafka [14] and its library Kafka Streams [15]. Figure 5 illustrates the high-level architecture of our prototype, where the shaded boxes represent the implemented modules. In this section, we first give a necessary background about Apache Kafka, and we next present the implementation details.

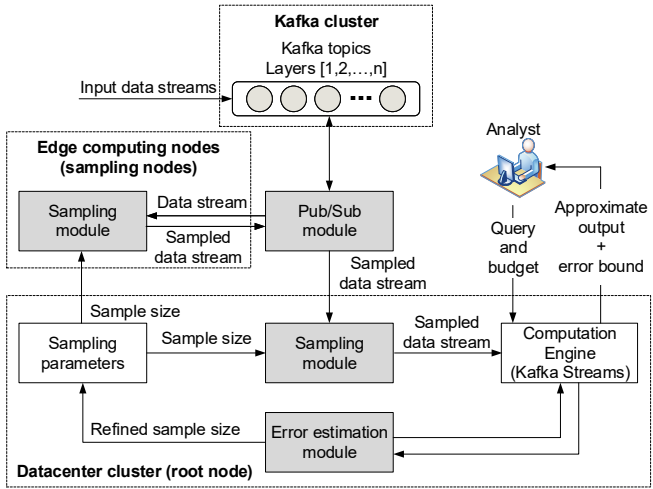

Fig. 5. APPROXIOT architecture.

## A. Background

Apache Kafka [14] is a widely used scalable fault-tolerant distributed pub/sub messaging platform. Kafka offers the reliable distributed queues called *topics* to receive input data streams. Stream analytics systems can subscribe these topics to retrieve and process data streams. We used Kafka to model the layers in the edge computing topology, where the input streams are pipelined across layers via pre-defined topics.

Recently, Kafka Streams [15] has been developed as a library on top of Kafka to offer a high-level dataflow API for stream processing. The key idea behind Kafka Streams is that it considers an input stream as an append-only data table (a log). Each arriving data item is considered as a row appended to the table. This design enables Kafka Streams to be a real-time stream processing engine, as opposed to the batched based stream processing systems (e.g., Spark Streaming [2]) that treat the input data stream as a sequence of micro-batches. Furthermore, since Kafka Streams is built on top of Kafka, it requires no additional cluster setup for a stream processing system (e.g., Apache Flink [16], Storm [17]). For these advantages, Kafka Streams is an excellent choice for our prototype implementation.

The Kafka Streams library supports two sets of APIs [15]: *(i)* High-Level Streams DSL (Domain Specific Language) API to build a processing topology (i.e., DAG dataflow) and *(ii)* Low-Level Processor API to create user-defined processors (a processor is an operator in the processing topology).

## B. APPROXIOT *Implementation Details*

At a high level (see Figure 5), the input data streams are ingested to a Kafka cluster.

**Edge computing nodes (sampling nodes).** A sampling node consumes an input stream from the Kafka cluster via the Pub/Sub module by subscribing to a pre-defined topic. Thereafter, the sampling module samples the input stream in an online manner using the proposed sampling algorithm (§III). Next, a producer is used to push the sampled data items to the next layer in the edge computing network topology using the Kafka topic of the next layer.

**Datacenter cluster (root node).** The root node receives the sampled data streams from the final layer of sampling nodes. First, it also makes use of the sampling module to take a sample of the input. Thereafter, the computation engine of Kafka Streams (High-Level Streams DSL processors) executes the input query over the sampled data stream to produce an approximate output. Finally, the error estimation module performs the error estimation mechanism (see §III-D) to provide a rigorous error bound for the approximate query result. In addition, in the case the error bound of the approximate result exceeds the desired budget of the user, an adaptive feedback mechanism is activated to refine the sampling parameters at all layers to improve the accuracy in subsequent runs. We next describe in detail the implemented modules.

**I: Pub/Sub module.** The Pub/Sub module ensures the communication between the edge computing layers. For that, we made use of the High-Level Streams DSL API to create the producer and consumer processors to send and retrieve data streams through a pre-defined topic corresponding to the layer.

**II: Sampling module.** The sampling module implements the algorithm described in §III. In particular, we implemented the algorithm in a user-defined processor (i.e., sampling processor) using the Low-Level API supported by Kafka. The sampling processor works as a normal processor in the Kafka computing topology to select input data items from the topics.

In addition, for the baseline comparison, we also implemented a simple random sampling (SRS) algorithm into a user-defined processor using the coin flip sampling algorithm [18].

**III: Error estimation module.** The error estimation module computes the error bounds for the approximate output, which is necessary for the user to interpret the accuracy of result. We used the Apache Common Math library [19] to implement the error estimation mechanism as described in §III-D.

## V. EVALUATION: MICROBENCHMARKS

In this section, we present the evaluation results of APPROXIOT using microbenchmarks. In the next section, we describe the evaluation results based on real-world datasets.

### A. Experimental Setup

**Cluster setup.** We deployed the APPROXIOT system using a cluster of 25 nodes. We used 15 nodes for the IoT deployment, each equipped with two dual-core Intel Xeon E3-1220 v3 processors and 4GB of RAM, running Ubuntu 14.04. In the deployment, we emulated a four-layer tree topology of an IoT infrastructure which contains 8 source nodes producing the input data stream, 4 nodes for the first edge computing layer, 2 nodes for the second edge computing layer, and one datacenter node (the root node). For the communication between the edge computing layers, we used a Kafka cluster using the 10 remaining nodes, each of which has 3-core Intel Xeon E5-2603 v3 processors and 8GB of RAM, running Ubuntu 14.04.

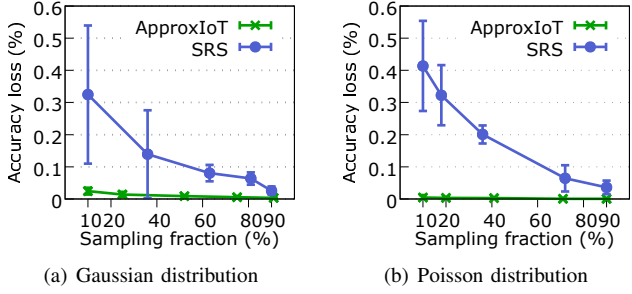
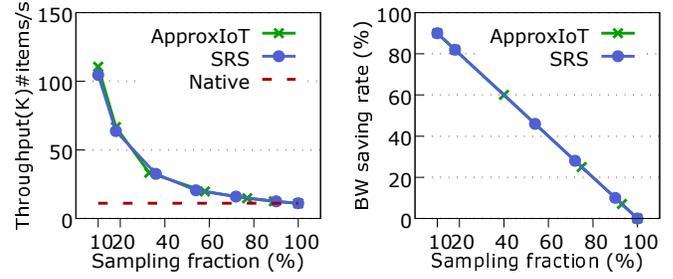

(a) Gaussian distribution  (b) Poisson distribution

Fig. 6. Accuracy loss vs sampling fraction. The accuracy loss of ApproxIoT is at most $0.035\%$ in (a) and $0.013\%$ in (b), both of which are smaller than the counterpart of SRS.

Fig. 7. Throughput vs sampling fraction.

Fig. 8. Bandwidth saving vs sampling fraction.

To emulate a WAN environment, we used the *tc* (traffic control) tool [20]. Based on the real measurements [21], the round-trip delay times between two adjacent layers are set to $20$ ms (between the source node and the first edge computing layer), $40$ ms (between the first layer and the second layer) and $80$ ms (between the second layer and the datacenter node). In the network, each link's capacity is $1$ Gbps. This WAN setting remains the same across all the experiments we conducted unless otherwise stated.

**Synthetic data stream.** We evaluated the performance of APPROXIOT using synthetic input data streams with two data distributions: Gaussian and Poisson. For the Gaussian distribution, we generated four types of input sub-streams: $\mathcal{A}$ ($\mu = 10, \sigma = 5$), $\mathcal{B}$ ($\mu = 1000, \sigma = 50$), $\mathcal{C}$ ($\mu = 10000, \sigma = 500$) and $\mathcal{D}$ ($\mu = 100000, \sigma = 5000$). For the Poisson distribution, we used four types of input sub-streams: $\mathcal{A}$ ($\lambda = 10$), $\mathcal{B}$ ($\lambda = 100$), $\mathcal{C}$ ($\lambda = 1000$) and $\mathcal{D}$ ($\lambda = 10000$).

**Metrics.** We evaluated the performance of APPROXIOT with the following three metrics: *(i) Throughput* defined as the number of data items processed per second; *(ii) Accuracy loss* defined as $|approx - exact|/exact$, where $approx$ and $exact$ denote the results produced by APPROXIOT and a native execution without sampling, respectively; and lastly, *(iii) Latency* defined as the end-to-end latency taken by a data item from the source until it is processed in the datacenter.

**Methodology.** We used the source nodes to produce and tune the rate of the input data streams such that the datacenter node in APPROXIOT was saturated. This input rate was used for three approaches: *(i)* APPROXIOT, *(ii) SRS-based system* employing Simple Random Sampling (in short, SRS), and *(iii) Native execution*. In the native execution approach, the input data streams are transferred from the source nodes all the way to the datacenter without any sampling at the edge nodes.

*B. Effect of Varying Sampling Fractions*

**Accuracy.** We first evaluate the accuracy loss of APPROXIOT and the SRS-based system. We use both Gaussian and Poisson distributions while we vary the sampling fractions.

Figure 6 shows that APPROXIOT achieves much higher accuracy than the SRS-based system for both datasets. In particular, when the sampling fraction is $10\%$, the accuracy of APPROXIOT is $10\times$ and $30\times$ higher than SRS's accuracy for Gaussian and Poisson datasets, respectively. This higher accuracy of APPROXIOT is because APPROXIOT ensures data items from each sub-stream are selected fairly by leveraging stratified sampling. Here, the absolute accuracy loss in SRS may look insignificant, but the estimation of SRS can be completely useless in the presence of a skewed distribution of arrival rates of the input streams, which we show in §V-E.

**Throughput.** We next evaluate the throughput of APPROXIOT in comparison with the SRS-based system.

Figure 7 depicts the throughput comparison between AP-PROXIOT and SRS. APPROXIOT achieves a similar throughput as SRS due to the fact that the proposed sampling mechanism, just like SRS, requires no synchronization between workers (CPU cores) to take samples from the input data stream. For instance, with the sampling fraction of $89\%$, the throughput of APPROXIOT is $12429$ items/s, and that of SRS is $12547$ items/s with the sampling fraction of $90\%$. Note that, as we perform sampling across different layers, we cannot ensure that two algorithms have the same sampling fraction.

Figure 7 also shows that both APPROXIOT and SRS have a similar throughput compared to the native execution even when the sampling fraction is $100\%$. APPROXIOT, SRS and the native execution achieve $11003$ items/s, $11046$ items/s and $11134$ items/s, respectively. This demonstrates the low overhead of our sampling mechanism.

**Network bandwidth.** In addition, sampling ensures that AP-PROXIOT (and SRS, too) significantly saves the network bandwidth between the computing layers as shown in Figure 8; the network resource is fully utilized in this case, so the sampling fraction of 10% means that our system only requires 10% of the total capacity (e.g., 100 Mbps out of 1 Gbps). Thus, even when the network resource is limited, APPROXIOT can function effectively.

**Latency.** We set the window size of APPROXIOT to one second. Figure 9 shows that APPROXIOT incurs a similar latency compared to the SRS-based system. In addition, when the sampling fraction of APPROXIOT is 10%, APPROXIOT achieves a $6\times$ speedup with respect to the native execution.

*C. Effect of Varying Window Sizes*

The previous window size of one second may look arbitrary. Thus, we evaluate the impact of varying window sizes on the latency of APPROXIOT. We set a fixed sampling fraction of 10% and measure the latency of the evaluated systems while we vary window sizes. Figure 10 shows the latency comparison between APPROXIOT and the SRS-based system. The latency of APPROXIOT increases as the window size increases whereas the latency of the SRS-based system remains the same. This is because the SRS-based system does not require a window for sampling the input streams in any of the edge computing layers. Therefore, like in any other window-based streaming systems [2], [16], the operators have to set small window sizes to meet the low latency requirement.

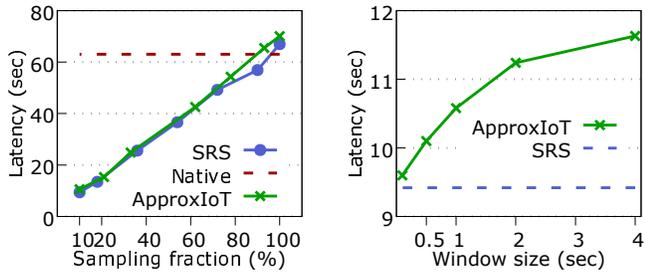

Fig. 9. Latency vs sampling fraction. APPROXIOT uses 1 second window.

Fig. 10. Latency vs window size. Sampling fraction is set to 10%.

*D. Effect of Fluctuating Input Rates of Sub-streams*

We next evaluate the impact of fluctuating rates of sub-streams on the accuracy of APPROXIOT. We keep the sampling fraction of 60% and measure the accuracy loss of APPROXIOT and the SRS-based system. Figures 11(a) and 11(b) present the accuracy loss of APPROXIOT and SRS with Gaussian distribution and Poisson distribution datasets. For these experiments, we create three different settings, in each of which four sub-streams $\mathcal{A}, \mathcal{B}, \mathcal{C}$ and $\mathcal{D}$ have different arrival rates. A setting is expressed as $(\mathcal{A} : \mathcal{B} : \mathcal{C} : \mathcal{D})$. For example, $(50k : 25k : 12.5k : 625)$ means that the input rates of sub-streams $\mathcal{A}, \mathcal{B}, \mathcal{C}$ and $\mathcal{D}$ are $50k$ items/s, $25k$ items/s, $12.5k$ items/s, and $625$ items/s, respectively.

Both figures show that the accuracy of these approaches improves proportionally to the input rate of the sub-stream $\mathcal{D}$ since data items of this sub-stream have significant values compared to other sub-streams. Across all settings, APPROXIOT achieves higher accuracy than the SRS-based system. For instance, under Setting1 in Figure 11(a), the accuracy loss of SRS-based system is $5.5\times$ higher than that of APPROXIOT; while under the same setting in Figure 11(b), the accuracy of APPROXIOT is $74\times$ higher than that of the SRS-based system. The higher accuracy of APPROXIOT against SRS is due to the similar reason that we already explained: the SRS-based system may overlook the sub-stream $\mathcal{D}$ in which there are only a few data items but their values are significant, whereas APPROXIOT is based on stratified sampling, and therefore, it captures all of the sub-streams well.

*E. Effect of Skew in Input Data Stream*

In this experiment, we analyze the effect of skew in the input data stream. We create a sub-stream that dominates the other sub-streams in terms of the number of data items. In particular, we generate an input data stream that consists of four sub-streams following a Poisson distribution, namely $\mathcal{A}$ ($\lambda = 10$), $\mathcal{B}$ ($\lambda = 100$), $\mathcal{C}$ ($\lambda = 1000$), and $\mathcal{D}$ ($\lambda = 10000000$). In this input data stream, the sub-stream $\mathcal{A}$ accounts for $80\%$ of all data items, whereas the sub-streams $\mathcal{B}, \mathcal{C}$ and $\mathcal{D}$ represent only $19.89\%$, $0.1\%$, and $0.01\%$, respectively.

Figure 11(c) shows that APPROXIOT achieves a significantly higher accuracy than the SRS-based system. With the sampling fraction of $10\%$, the accuracy of APPROXIOT is $2600\times$ higher than the accuracy of SRS-based system. The reason for this is that APPROXIOT considers each sub-stream fairly — none of them is ignored when samples are taken. Meanwhile, the SRS-based system may not yield sufficient numbers of data items for each sub-stream. Interestingly, as highlighted in Figure 11(c), the SRS-based system may overestimate the sum of the input data stream since it by chance mainly considers sub-stream $\mathcal{D}$ and ignores others (see evaluation results with the sampling fraction of $10\%$).

## VI. EVALUATION: REAL-WORLD DATASETS

In this section, we evaluate APPROXIOT using two real-world datasets: *(i)* New York taxi ride and *(ii)* Brasov pollution dataset. We used the same cluster setup as described in §V-A.

*A. New York Taxi Ride Dataset*

**Dataset.** The NYC taxi ride dataset has been published at the DEBS 2015 Grand Challenge [22]. This dataset consists of the ride information of $10,000$ taxies in New York City in 2013. We used the dataset from January 2013.

**Query.** We performed the following query: *What is the total payment for taxi fares in NYC at each time window?*

**Results.** Figure 12(a) shows that the accuracy of APPROXIOT improves with the increase of sampling fraction. With the sampling fraction of $10\%$, the accuracy loss of APPROXIOT is $0.1\%$, whereas with the sampling fraction of $47\%$, the accuracy loss is only $0.04\%$. In addition, we measure the throughput of APPROXIOT with varying sampling fractions. Figure 12(b) depicts that the throughput of APPROXIOT reduces when the sampling fraction increases. With the sampling fraction of $10\%$, the throughput of APPROXIOT is 122,199 items/sec, which is roughly $10\%$ higher than the native execution.

*B. Brasov Pollution Dataset*

**Dataset.** The Brasov pollution dataset [23] consists of the pollution measurements (e.g., air quality index) in Brasov, Romania from August 2014 to October 2014. Each sensor provides a measurement result every $5$ minutes.

**Query.** We performed the following query: *What is the total pollution values of particulate matter, carbon monoxide, sulfur dioxide and nitrogen dioxide in every time window?*

**Results.** Figure 12(a) depicts the accuracy loss of APPROXIOT in processing the pollution dataset with varying sampling fractions. With the sampling fractions of $10\%$ and $40\%$, the accuracy loss of APPROXIOT are $0.07\%$ and $0.02\%$, respectively. The accuracy loss in processing this dataset has a similar but lower curve as for the NYC taxi ride dataset. The reason is that the values of data items in Brasov pollution dataset are more stable than in NYC tax ride dataset.

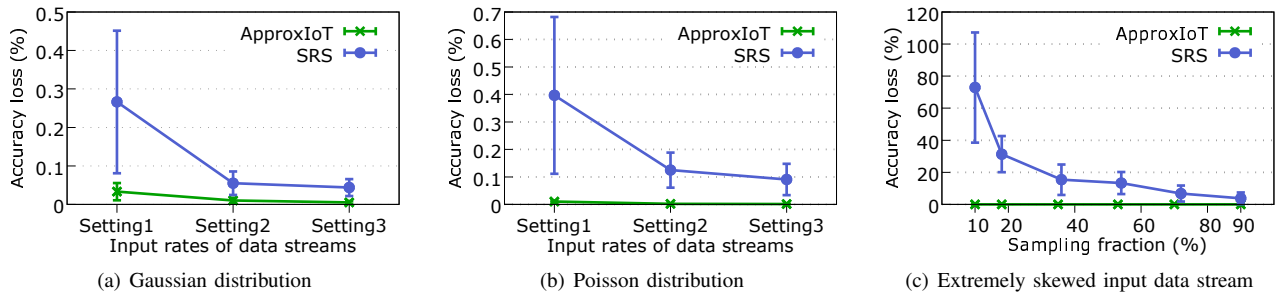

(a) Gaussian distribution  (b) Poisson distribution  (c) Extremely skewed input data stream

Fig. 11. The accuracy comparison between APPROXIOT and the SRS-based system with different arrival rates of sub-streams. For (a) and (b), the arrival rates (items/sec) of the four input sub-streams $\mathcal{A}$, $\mathcal{B}$, $\mathcal{C}$, and $\mathcal{D}$ are the following: Setting1: $(50k : 25k : 12.5k : 625)$, Setting2: $(25k : 25k : 25k : 25k)$ and Setting3: $(625 : 12.5k : 25k : 50k)$. For (c), Poisson distribution is used; $\mathcal{A}$, $\mathcal{B}$, $\mathcal{C}$ and $\mathcal{D}$ have $\lambda = 10, 100, 1000$ and $10000000$, respectively; the sub-stream $\mathcal{A}$ accounts for 80% of all data items while the sub-streams $\mathcal{B}$, $\mathcal{C}$ and $\mathcal{D}$ account for only 19.89%, 0.1%, and 0.01%, respectively. The average accuracy loss of APPROXIOT is at most 0.056% in (a), 0.014% in (b) and 0.035% in (c).

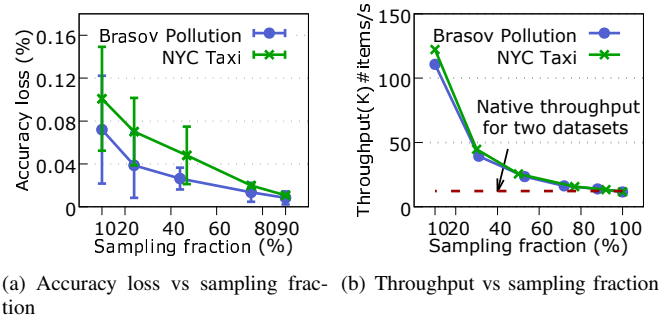

(a) Accuracy loss vs sampling fraction  (b) Throughput vs sampling fraction

Fig. 12. The accuracy loss and throughput of APPROXIOT in processing the two real-world datasets. The flat line in (b) shows the throughput of the native approach for processing the two datasets; only one line is presented because there is a marginal difference between processing the two datasets.

Figure 12(b) presents the throughput of APPROXIOT with different sampling fractions. With the sampling fraction of 10%, APPROXIOT achieves a $9\times$ higher throughput than the native execution. The throughputs of processing both the NYC taxi ride dataset and the pollution dataset are similar.

## VII. RELATED WORK

With the ability to enable a systematic trade-off between accuracy and efficiency, approximate computing has been explored in the context of distibuted data analytics [24], [25], [26], [27], [28], [9], [?]. In this context, sampling-based techniques are properly the most widely used for approximate data analytics [24], [25], [26]. These systems show that it is possible to leverage the benefits of approximate computing in the distributed big data analytics settings. Unfortunately, these systems are mainly targeted towards batch processing, where the input data remains unchanged during the course of sampling. Therefore, these systems cannot cater to stream analytics, which requires real-time processing of data streams.

To overcome this limitation, IncApprox [27], StreamApprox [9], [29], and ApproxJoin [30] have been proposed for approximate stream analytics. IncApprox introduces an online "biased sampling" algorithm that uses self-adjusting computation [31] to produce incrementally updated approximate results [32], [33], [34], [35]. Meanwhile, StreamApprox handles the fluctuation of input streams by using an online adaptive stratified sampling algorithm. These systems demonstrate that it's also possible to trade the output quality for efficiency in stream processing. ApproxJoin proposed an approximate join mechanism for distributed data analytics. Unfortunately, these systems target processing input data streams within a centralized datacenter, where the online sampling is carried out at a centralized stream aggregator. In APPROXIOT, we designed a distributed online sampling algorithm for the IoT setting, where the sampling is carried out in a truly distributed fashion at multiple levels using the edge computing resources.

Recently, in the context of IoT, *edge computing* has emerged as a promising solution to reduce latency in data analytics systems [36], [37]. In edge computing, a part of computation and storage are performed at the Internet's edge closer to IoT devices or sensors. By moving either whole or partial computation to the edge, edge computing allows to achieve not only low latency but also significant reduction in bandwidth consumption [37]. Several works deploy sampling and filtering mechanisms at sources (sensor nodes) to further optimize communication costs [38], [39]. However, the proposed sampling mechanisms in these works are "snapshot sampling" techniques which are used to take input data stream every certain time interval. PrivApprox [28], [40] proposed a marriage of approximate computing based on sampling with the randomized response for improved performance and users' privacy. As opposed, in APPROXIOT, we leverage sampling-based techniques at the edge to further reduce the latency and bandwidth consumption in processing large-scale IoT data. In detail, we design an online adaptive random sampling algorithm, and perform it not only at the root node, but also at all layers of the computing topology.

Finally, it is worth to mention that there has been a surge of research in geo-distributed data analytics in multi-datacenters [41], [42], [43], [44], [45], [46]. However, these system focus on improving the performance for batch processing in the context of data centers, and are not designed for edge computing. In APPROXIOT, we design an approximation technique for real-time stream analytics in a geo-distributed edge computing.

## VIII. CONCLUSION

The unprecedentedly huge volume of data in the IoT era presents both opportunities and challenges for building data-driven intelligent services. The current centralized computing model cannot cope with low-latency requirement in many online services, and it is also a wasteful computing medium in terms of networking, computing, and storage infrastructure for handling IoT-driven data streams across the globe. In this paper, we explored a radically different approach that exploits approximate computing paradigm for a globally distributed IoT environment. We designed and implemented APPROXIOT, a stream analytics system for IoT that achieves efficient resource utilization, and also adapts to the varying requirements of analytics applications and constraints in the underlying computing/networking infrastructure. The nodes in the system run a weighted hierarchical sampling algorithm independently without any cross-node coordination, which facilitates parallelization, thereby making APPROXIOT scalable. Our evaluation with synthetic and real-world datasets demonstrates that APPROXIOT achieves $1.3\times$—$9.9\times$ higher throughput than the native stream analytics execution and $3.3\times$—$8.8\times$ higher accuracy than a simple random sampling scheme under the varying sampling fractions of $80\%$ to $10\%$.

**Limitations and future work.** While APPROXIOT approach is quite useful to achieve desired properties, our current system implementation has the following limitations.

First, APPROXIOT currently supports only *approximate linear queries*. We plan to extend the system to support more complex queries [47], [26] such as joins, top-$k$, etc., as part of the future work.

Second, our current implementation relies on manual adjustment of user's query budget to the required sampling parameters. As part of the future work, we plan to implement an automated cost function to tune the sampling parameters for the required system performance and resource utilization.

Lastly, we have evaluated APPROXIOT using a small testbed. As part of the future work, we plan to extend our system evaluation via deploying APPROXIOT over Azure Stream Analytics [48] to further evaluate the performance of our system in a real IoT infrastructure.

The source code of APPROXIOT is publicly available: https://ApproxIoT.github.io/ApproxIoT/